\documentclass{mn2e}
\usepackage{times,graphicx,astrobib,amssymb,aas_macros}

\newcommand{\be}{\begin{equation}}
\newcommand{\e}{\end{equation}}
\newcommand{\bear}{\begin{eqnarray}}
\newcommand{\ear}{\end{eqnarray}}
\newcommand{\nline}{\nonumber \\}
\newcommand{\f}{\frac}
\newcommand{\de}{{\rm d}}
\newcommand{\del}{\partial}

\newcommand{\matrixsymbol}{\sf}

\begin{document}

\title[Reionization constraints using PCA]{Reionization constraints using Principal Component Analysis}
\author[Mitra, Choudhury \& Ferrara]
{Sourav Mitra$^1$\thanks{E-mail: smitra@hri.res.in},~
T. Roy Choudhury$^1$\thanks{E-mail: tirth@hri.res.in}~
and
Andrea Ferrara$^2$\thanks{E-mail: andrea.ferrara@sns.it}
\\
$^1$Harish-Chandra Research Institute, Chhatnag Road, Jhusi, Allahabad 211019, India\\
$^2$Scuola Normale Superiore, Piazza dei Cavalieri 7, 56126 Pisa, Italy
} 

\maketitle

\date{\today}

\begin{abstract}
Using a semi-analytical model developed by \citeN{2005MNRAS.361..577C}
we study the observational constraints on reionization via a
principal component analysis (PCA).
Assuming that reionization at $z > 6$ is primarily driven by stellar
sources, we decompose the
unknown function $N_{\rm ion}(z)$, representing the number of photons in the IGM
per baryon in collapsed objects, into its principal components and
constrain the latter using the photoionization rate, 
$\Gamma_{\rm PI}$, obtained from Ly$\alpha$ forest Gunn-Peterson
optical depth, the WMAP7 electron scattering optical depth $\tau_{\rm el}$
and the redshift distribution of Lyman-limit systems $\de N_{\rm LL}/\de z$ 
at $z \sim 3.5$. The main findings of our analysis are: (i) It is sufficient to 
model $N_{\rm ion}(z)$ over the redshift range $2 < z < 14$ using 5 parameters 
to extract the maximum information contained within the data. (ii) All quantities 
related to reionization can be severely constrained for $z < 6$ because of a large number of data points whereas constraints at $z > 6$ are relatively loose. (iii) 
The weak constraints on $N_{\rm ion}(z)$ at $z > 6$ do not allow to disentangle different feedback models with present data. There is a clear indication that $N_{\rm ion}(z)$ must increase at $z > 6$, thus ruling out reionization by a single stellar population 
with non-evolving IMF, and/or star-forming efficiency, and/or photon escape fraction. The data allows for non-monotonic $N_{\rm ion}(z)$ which may contain sharp features around $z \sim 7$. (iv) The PCA implies that reionization must be 99\% completed between $5.8 < z < 10.3$ (95\% confidence level) and is expected to be 50\% complete at $z \approx  9.5-12$. With future data sets, like those obtained by {\it Planck}, the $z > 6$ constraints will be significantly improved.
\end{abstract}

\section{Introduction}

The importance of studying hydrogen reionization at high redshifts
lies in the fact that it is tightly coupled to properties of
first luminous sources and subsequent galaxy formation
(for reviews, see, \citeNP{2001ARA&A..39...19L,2001PhR...349..125B,2006astro.ph..3149C,2009arXiv0904.4596C}). In recent years,
studies in reionization have been boosted by (i) the availability of a
wide range of data sets and (ii) the expectation that the volume of data would
increase rapidly over the next few years (for reviews, see 
\citeNP{2006PhR...433..181F,2006ARA&A..44..415F}).
Given such a large amount of data,
it is important to develop theoretical and statistical methods so that
maximum information can be extracted.

Theoretically, reionization is modelled either semi-analytically 
or by numerical simulations.
Unfortunately, the physical processes
relevant to reionization are
so complex that neither of the two approaches can capture the overall 
picture entirely. The simulations are indispensable for
understanding detailed spatial distribution of ionized regions and
topology of reionization. However, if one is interested in 
the evolution of globally-averaged quantities, then 
semi-analytical models prove to be very useful in providing insights. 
The main reason for this is that these models can probe a wide range
of parameter space which can be quite large depending on
our ignorance of the different processes.

At present, our understanding of reionization is that it is primarily
driven by ultra-violet radiation from 
stellar sources forming within galaxies. 
The major uncertainty in modelling reionization
is to model the star-formation history and transfer of radiation from the
galaxies to the intergalactic medium (IGM) which is usually
parameterized through $N_{\rm ion}$, the number of photons entering 
the IGM per baryon in collapsed objects. This parameter, in principle,
has a dependence on $z$ which can arise from evolution of 
star-forming efficiency, fraction of photons escaping from the host halo
and chemical and radiative feedback processes. Note that this parameter
remains uncertain even in numerical simulations, hence the
semi-analytical models can become handy in studying a wide range 
of parameter values and the corresponding agreement with data sets. 
In analytical studies, $N_{\rm ion}(z)$ is either taken to be a 
piecewise constant
function \cite{2003ApJ...586..693W,2005MNRAS.361..577C} 
or parameterized using some known functions 
\cite{2003ApJ...599..759C,2010MNRAS.408...57P} or
modelled using a physically-motivated
prescription \cite{2006MNRAS.371L..55C}. 
In particular, a model involving metal-free and normal
stars with some prescription for radiative and chemical feedback 
can match a wide range of observations \cite{2006MNRAS.371L..55C,2006MNRAS.370.1401G} and possibly make prediction regarding
search for reionization sources by future experiments \cite{2007MNRAS.380L...6C}.

However, the fact remains that many of the physical processes involved
in modelling $N_{\rm ion}$ are still uncertain. 
Given this, it
is worthwhile doing a detailed probe of the parameter space and determine
the range of reionization histories that are allowed by the data. 
In other words, rather than working out the uncertain physics, one can ask
the question as to what are the forms of $N_{\rm ion}(z)$ implied by the data 
itself.  It is expected
that in near future, with more data sets becoming available, the allowed 
range in the forms of $N_{\rm ion}(z)$ would be severely constrained, 
thus telling us exactly how reionization
occurred. 
Now, it is obvious that the constraints on $N_{\rm ion}(z)$ will 
not be same for all redshifts, points where there are more and better data
available, the constraint would be more tight. Similarly, since we deal
with a heterogeneous set of data, it is expected that the constraints
would depend on the nature of data used.
It is thus important to know which aspects of reionization
history can be constrained by what kind of data sets. A method which is
ideally suited to tackle this problem is to use the principal component analysis
(PCA); this is a technique to compute the most meaningful basis to re-express 
the unknown parameter set and the hope is that this 
new basis will reveal hidden detailed statistical structure.

In this work, we make a preliminary attempt to constrain $N_{\rm ion}(z)$ using
PCA and hence estimate the uncertainties in 
the reionization history. 
The main objective of the work would be to find out the widest possible
range in reionization histories allowed by the different data sets.

Throughout the paper,  we
assume a flat Universe with cosmological parameters  given by the 
WMAP7 best-fit values: $\Omega_m =
0.27$, $\Omega_{\Lambda} = 1 - \Omega_m$, $\Omega_b h^2 = 0.023$, and
$h=0.71$.  The  parameters defining the linear dark  matter power
spectrum we use are $\sigma_8=0.8$, $ n_s=0.96$, $\de n_s/\de \ln k =0$
\cite{2010arXiv1001.4635L}.

\section{Semi-analytical model of reionization}
\label{sec:cfmodel}

\subsection{Features of the model}

The semi-analytical model used in this work is based on 
\citeN{2005MNRAS.361..577C} and 
\citeN{2006MNRAS.371L..55C}.
Let us first summarize the main features of the model 
alongwith the modifications made in this work:

\begin{itemize}

\item The model accounts for IGM inhomogeneities by adopting a lognormal distribution according to the method outlined in \citeN{2000ApJ...530....1M}; 
reionization is said to be complete once all the low-density regions (say, with overdensities $\Delta < \Delta_{\rm crit} \sim 60$) are ionized. 
The mean free path of photons is thus determined essentially by the distribution of high density regions:
\be
\lambda_{\rm mfp}(z) = \f{\lambda_0}{[1 - F_V(z)]^{2/3}}
\label{eq:lambda_0}
\e
where $F_V$ is the volume fraction of ionized regions and $\lambda_0$ is a normalization parameter. In our earlier works, the value of this parameter was fixed 
by comparing with low redshift observations while in this work, we 
treat it as a free parameter.
  We follow the ionization and thermal histories 
of neutral, HII and HeIII regions simultaneously and self-consistently, treating the IGM as a multi-phase medium.

\item The model assumes that reionization is driven by stellar sources.
The stellar sources can further be divided into two classes, namely,
(i) metal-free (i.e. PopIII) stars having a Salpeter IMF in the mass 
range $1 - 100 M_{\odot}$: they dominate the photoionization rate at high redshifts; (ii) PopII stars with sub-solar metallicities 
also having a Salpeter IMF in the mass range $1 - 100 M_{\odot}$. 

\item Reionization by UV sources is accompanied by photo-heating of the gas, which can result in a suppression of star formation in low-mass 
haloes. We compute such (radiative) feedback self-consistently from the evolution of the thermal properties of the IGM.

\item Furthermore the \textit{chemical feedback} including PopIII$\rightarrow$PopII transition is implemented using merger-tree based genetic 
approach \citeN{2006MNRAS.369..825S}. Under this approach, it is assumed that
if a given star-forming halo has a progenitor which formed PopIII stars, 
then the halo under consideration is enriched and cannot 
form PopIII stars. In this work, we introduce an analytical formula for the transition from PopIII to PopII phase using the conditional probability of 
Press-Schechter mass function \cite{1993MNRAS.262..627L}. 
The probability that a halo of mass $M$ at $z$ never had a progenitor in the mass range 
$[M_{\rm min}(z),M+M_{\rm res}]$ is given by 
\begin{equation}
  f_{\rm III}(M,z) = \frac{2}{\pi} \tan^{-1} 
\left[\frac{\sigma(M+M_{\rm res})-\sigma(M)}
{\sigma(M_{\rm min}(z))-\sigma(M+M_{\rm res})}\right],
\end{equation}
where $M_{\rm min}$ is the minimum mass of haloes which are able to
form stars and $M_{\rm res}$ represents the minimum increase in mass 
(either by accretion or by merger) of an object
so that it may be identified as a new halo.
The fraction of collapsed haloes 
which are able to form PopII and PopIII stars at redshift $z$ are given by the
following relations:
\bear
f_{\rm coll,II}(z)  &\!\!\!\!=\!\!\!\!& 
\f{1}{\bar{\rho}_m} \int_{M_{\rm min}(z)}^{\infty} \de M~
[1 - f_{\rm III}(M,z)] M \f{\del n(M,z)}{\del M},
\nline
f_{\rm coll,III}(z) &\!\!\!\!=\!\!\!\!& 
\f{1}{\bar{\rho}_m} \int_{M_{\rm min}(z)}^{\infty} \de M~
f_{\rm III}(M,z) M \f{\del n(M,z)}{\del M}.
\ear
with $f_{\rm coll,II}(z) + f_{\rm coll,III}(z) = f_{\rm coll}(z)$.
The quantity $\bar{\rho}_m$ is the comoving density of dark matter
and $\del n/\del M$ is number density of collapsed objects per unit comoving 
volume per unit mass range \cite{1974ApJ...187..425P}.

\item
Given the collapsed fraction, this model calculates the production rate of ionizing photons in the IGM as
\be
\dot{n}_{\rm ph}(z) = n_b
\left[N_{\rm ion, II} \f{\de f_{\rm coll,II}}{\de t}
+ N_{\rm ion, III} \f{\de f_{\rm coll,III}}{\de t}
\right]
\e
where $n_b$ is the total baryonic number density in the IGM 
and $N_{\rm ion, II} (N_{\rm ion, III})$ is the number of photons 
from PopII (PopIII) stars entering the IGM 
per baryon in collapsed objects. The parameter $N_{\rm ion}$ can actually
be written as a 
combination of various other parameters:
\be
N_{\rm ion} \equiv \epsilon_* f_{\rm esc} m_p \int_{\nu_{\rm HI}}^{\infty} \de \nu
\left[\f{\de N_{\nu}}{\de M_*}\right]
\equiv \epsilon m_p \int_{\nu_{\rm HI}}^{\infty} \de \nu
\left[\f{\de N_{\nu}}{\de M_*}\right],
\e
where $\epsilon_*$ denotes the star-forming efficiency 
(fraction of baryons within
collapsed haloes going into stars), $f_{\rm esc}$ is the fraction of photons
escaping into the IGM, $\left[\de N_{\nu}/\de M_*\right]$ gives the number of photons emitted per  
frequency range per unit mass of stars (which depends on the stellar IMF and
the corresponding stellar spectrum) and 
$\epsilon \equiv \epsilon_* f_{\rm esc}$. For PopII stars
with sub-solar metallicities 
having a Salpeter IMF in the mass range $1 - 100 M_{\odot}$, we get
$N_{\rm ion, II} \approx 3200 \epsilon_{\rm II}$, while for 
PopIII stars having a Salpeter IMF in the mass 
range $1 - 100 M_{\odot}$, we get $N_{\rm ion, III} \approx 35000 \epsilon_{\rm III}$.

In this Section, we take $\epsilon_{\rm II}, \epsilon_{\rm III}$
(or, equivalently $N_{\rm ion, II}, N_{\rm ion, III}$) to be independent of 
$z$ and $M$, which
implies that the star-forming efficiencies and the escape fractions do
not depend on the mass of the star-forming halo and also do not
evolve. However, note that the effective $N_{\rm ion}$ (which is the
appropriately weighted average of  $N_{\rm ion, II}$ and  $N_{\rm ion, III}$)
evolves with $z$
\be
N_{\rm ion}(z) = \f{N_{\rm ion, II} \f{\de f_{\rm coll,II}}{\de t}
+ N_{\rm ion, III} \f{\de f_{\rm coll,III}}{\de t}}
{\f{\de f_{\rm coll,II}}{\de t} + \f{\de f_{\rm coll,III}}{\de t}}
\label{eq:nion_eff}
\e
At high redshifts, we expect $\de f_{\rm coll,II}/\de t \to 0$, hence
$N_{\rm ion}(z) \to N_{\rm ion, III}$, and similarly at low redshifts where
chemical enrichment is widespread, we have
$N_{\rm ion}(z) \to N_{\rm ion, II}$.

\item We also include the contribution of quasars based on their
observed luminosity function at $z < 6$ \cite{2007ApJ...654..731H}; 
we assume that 
they have negligible effects on IGM at higher redshifts.
They are significant sources of photons at $z\lesssim 4$ and are
particularly relevant for studying helium reionization.

\item The free parameters for this analysis would be 
$\epsilon_{\rm II}, \epsilon_{\rm III}$ (or, equivalently 
$N_{\rm ion, II}, N_{\rm ion, III}$) and $\lambda_0$, the normalization which 
determines the mean free path of photons. 

\item Usually, the model is constrained by comparing with a variety
of observational data, namely, (i) redshift evolution of Lyman-limit absorption 
systems (LLS), (ii)  IGM Ly$\alpha$ and Ly$\beta$ optical depths, (iii) electron scattering optical depth, 
(iv) temperature of the mean intergalactic gas, and (v)  cosmic star formation history.
However, most of the constraints on the model come from a subset of the
above data sets.
In this work, we would like to carry out a detailed 
likelihood analysis of the parameters. Hence to keep the analysis simple, 
the likelihood analysis is done using only three particular data sets which 
are discussed as follows:

(i) We use estimates for the photoionization rates 
$\Gamma_{\rm PI}$ obtained using Ly$\alpha$ forest Gunn-Peterson
optical depth observations and a large set of hydrodynamical simulations
\cite{2007MNRAS.382..325B}. The error-bars in these data points take into
account the uncertainties in the thermal state of the IGM in addition to
the observational errors in the Ly$\alpha$ optical depth. The data points 
have a mild dependence on the cosmological parameters which has been
taken into account in this work. We also find that although the
error-bars on $\Gamma_{\rm PI}$ are highly asymmetric, 
those on $\log(\Gamma_{\rm PI})$ are relatively symmetric; hence we 
use values of $\log(\Gamma_{\rm PI})$ and the corresponding errors in our
likelihood analysis.
The photoionization rate can be
obtained in our model from $\dot{n}_{\rm ph}(z)$ 
using the relation
\be
\Gamma_{\rm PI}(z) = 
(1 + z)^3 \int_{\nu_{\rm HI}}^{\infty} \de \nu ~ \lambda_{\rm mfp}(z;\nu) 
\dot{n}_{\rm ph}(z; \nu) \sigma_H(\nu) 
\e
where $\nu$ the frequency of radiation, $\nu_{\rm HI}$ is the 
threshold frequency for photoionization of hydrogen and 
$\sigma_H(\nu)$ is the photoionization cross section of hydrogen.

(ii) The second set of observations we have used corresponds to the WMAP7
data on electron scattering optical depth $\tau_{\rm el}$
\cite{2010arXiv1001.4635L}. The reported value of this quantity 
depends on the background cosmological model used. In this work, we
restrict ourselves to the flat CDM universe with a cosmological constant
and use the corresponding constraints on $\tau_{\rm el}$. 
Also, the $\tau_{\rm el}$ constraint is treated as a single data point
which should be thought as a simplification
because CMB polarization 
observations are, in principle, sensitive to the shape of the 
reionization history \cite{2008MNRAS.385..404B}. 
However, we have checked and found that the range of reionization histories
considered in this paper would hardly make any difference to the 
currently observed large angular scale polarization anisotropies 
other than the value of $\tau_{\rm el}$.
The quantity $\tau_{\rm el}$ can be obtained from our model 
given the global reionization history, in particular the
comoving density of free electrons $n_e(z)$:
\be
\tau_{\rm el}(z) = \sigma_T c \int_0^{z[t]} \de t ~ n_{e} ~ 
(1+z)^3 
\e
where $\sigma_T$ is the Thomson scattering cross section.

(iii) Finally, we use the redshift distribution of
LLS $\de N_{\rm LL}/\de z$ 
at $z \sim 3.5$ \cite{2010ApJ...718..392P}.\footnote{We did not include the more recent measurements of $\de N_{\rm LL}/\de z$ by \citeN{2010ApJ...721.1448S}
because the values are systematically larger than the ones quoted in 
\citeN{2010ApJ...718..392P} at $z \sim 3.5$; inclusion of both the data sets would lead to a bad fit for the model. The \citeN{2010ApJ...721.1448S} set has a data point at 
$z \sim 6$ which is not present in other data sets, however the present error-bar on that particular point is relatively large and hence excluding it does not affect our constraints significantly.} 
The data points are obtained using a large sample of QSO spectra which
results in extremely small statistical errors. However, there are
various systematic effects arising from effects like the incidence of
proximate LLS and uncertainties in the continuum. Usually, these
effects contribute to about 10--20\% uncertainty in the data points.
The quantity $\de N_{\rm LL}/\de z$ can be calculated in our model
from the mean free path:
\be
\f{\de N_{\rm LL}}{\de z}
= \f{c}{\sqrt{\pi}~\lambda_{\rm mfp}(z) H(z) (1+z)}
\e
Note that inclusion of the Lyman-limit systems
in the analysis is crucial for constraining the parameter $\lambda_0$. 

The likelihood function used in our calculations is given by
\be
L \propto \exp (-{\cal L})
\e
where ${\cal L}$ is the negative of the log-likelihood. It is estimated 
using the relation
\be
{\cal L} = \f{1}{2}\sum_{\alpha=1}^{n_{\rm obs}} \left[
\f{{\cal G}_{\alpha}^{\rm obs} - {\cal G}_{\alpha}^{\rm th}}{\sigma_{\alpha}}
\right]^2
\e
where  ${\cal G}_{\alpha}$ represents the 
set of $n_{\rm obs}$ observational data points described above, i.e.,  
${\cal G}_{\alpha} = \{\log(\Gamma_{\rm PI}), \tau_{\rm el}, \de N_{\rm LL}/\de z\}$
and $\sigma_{\alpha}$ are the corresponding observational error-bars.
We constrain the free parameters by maximizing the likelihood function.
We impose a prior
such that reionization should be complete by $z = 5.8$, otherwise it will
not match that Ly$\alpha$ and Ly$\beta$ forest transmitted flux data.

\end{itemize}

\subsection{Reionization Constraints}
\label{sec:reion_constraints}

\begin{figure*}
\includegraphics[height=\textwidth, angle=270]{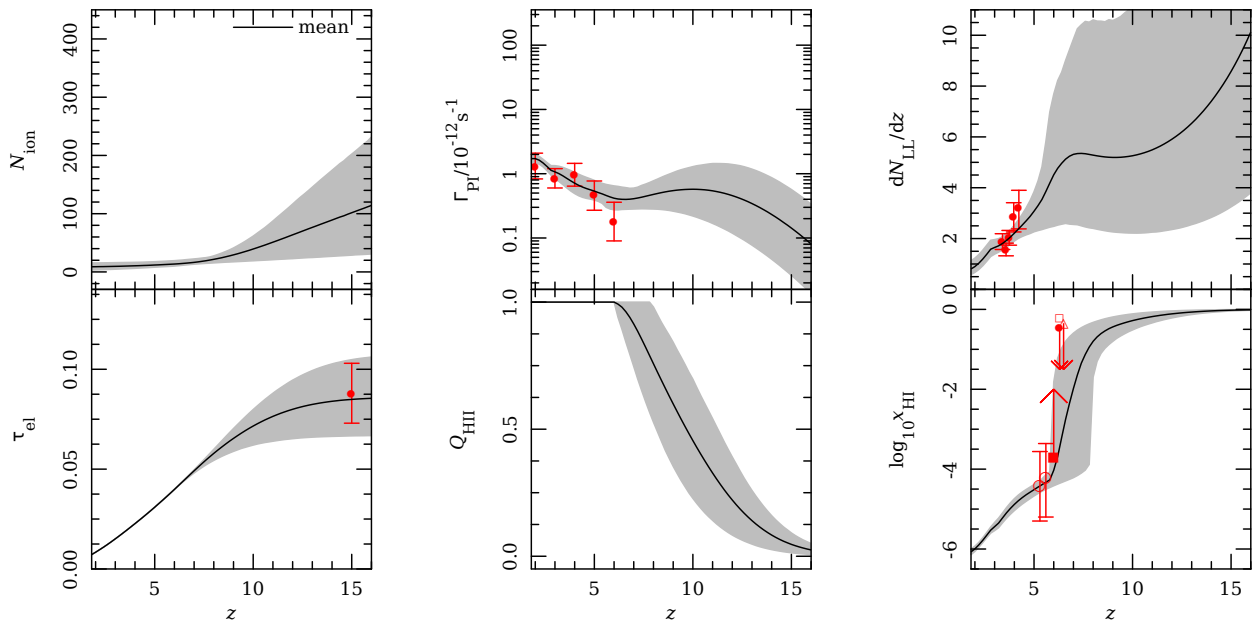}
\caption{The marginalized posteriori distribution of various 
quantities related to
reionization history for a model with chemical feedback 
(Choudhury \& Ferrara 2006).
The solid lines correspond to the model described by mean values of the
parameters
while the shaded regions correspond to 2-$\sigma$ limits.
The points with error-bars denote the observational data points.
{\it Top-left:} the evolution of 
the effective $N_{\rm ion}(z)$;
{\it Top-middle:} 
the hydrogen photoionization rate $\Gamma_{\rm PI}(z)$ 
alongwith the constraints from Bolton \& Haehnelt (2007);
{\it Top-right:} 
the LLS distribution $\de N_{\rm LL}/\de z$ with data points from Prochaska, O'Meara \& Worseck (2010); 
{\it Bottom-left:} the electron scattering optical depth $\tau_{\rm el}$ with
the WMAP7 constraint (Larson et al. 2010);
{\it Bottom-middle:} 
the volume filling factor of HII regions $Q_{\rm HII}(z)$;
{\it Bottom-right:}
the global neutral hydrogen fraction $x_{\rm HI}(z)$ with observational
limits from QSO absorption lines (Fan et al. 2006; filled square), 
Ly$\alpha$ emitter luminosity function (Kashikawa et al. 2006; open triangle) 
and GRB spectrum analysis (Totani et al 2006; open square). 
Also shown
are the constraints using dark gap statistics on QSO spectra 
(Gallerani et al 2008a; open circles) and GRB spectra (Gallerani et al. 2008b;
filled circle).
}
\label{fig:threep_posteriori}
\end{figure*}
\nocite{2006MNRAS.371L..55C,2007MNRAS.382..325B,2010ApJ...718..392P,2010arXiv1001.4635L,2006AJ....131.1203F,2006PASJ...58..485T,2006ApJ...648....7K,2008MNRAS.386..359G,2008MNRAS.388L..84G}

The results of our likelihood analysis using the reionization model described 
above are summarized in Table \ref{tab:threep_likelihood}. 
The evolution of various quantities for models which are allowed
within 95\% confidence limit is 
shown in Figure \ref{fig:threep_posteriori}.

\begin{table}
\begin{tabular}{l|l|l}
Parameters & Mean value & 95\% confidence limits\\
\hline
$\epsilon_{\rm II}$ & $0.003$ & $[0.001, 0.005]$\\
$\epsilon_{\rm III}$ & $0.020$ & $[0.000, 0.043]$\\
$\lambda_0$ & $5.310$ & $[2.317, 9.474]$\\
\hline
$z(Q_{\rm HII}=0.5)$ & $9.661$ & $[7.894, 11.590]$\\
$z(Q_{\rm HII}=0.99)$ & $6.762$ & $[5.800, 7.819]$\\
\hline
\end{tabular}
\caption{The marginalized posterior probabilities
 with 95\% C.L. errors of all free parameters (top three parameters) and derived
parameters (from the fourth parameter down) for the reionization model
with PopII and PopIII stars.
}
\label{tab:threep_likelihood}
\end{table}

The top-left panel of the figure shows the evolution of the effective
$N_{\rm ion}$ as given by equation (\ref{eq:nion_eff}). One can see that the
quantity attains a constant value $\approx 10$ at $z < 6$ which is a
consequence of the fact that the photon emissivity at those epochs are
purely determined by PopII stars. However at higher redshifts, the
value of $N_{\rm ion}$ increases with $z$ because of the presence of 
PopIII stars. It is clear that the data cannot be fitted with PopII stars
with constant $N_{\rm ion, II}$ alone, one requires a rise in $N_{\rm ion}$ at
higher redshifts. For the kind of chemical feedback employed in the model, the
rise is rather smooth and gradual.

The mean values of parameters quoted in Table \ref{tab:threep_likelihood}
are similar to the best-fit model described in \citeN{2006MNRAS.371L..55C} and 
hence the corresponding
reionization history is similar to those described in the same paper. This can
be readily verified from Figure \ref{fig:threep_posteriori} where we see that
reionization starts around $z \approx 15$ driven by PopIII stars, 
and it is 90 per cent complete by $z \approx 7.5$.  
After a rapid initial phase, the growth of the
volume filled by ionized regions slows down at $z \lesssim 10$ due to the 
combined action of chemical and radiative feedback,  
making reionization a considerably extended process completing only at 
$z \approx 6$. 
We refer the reader 
to our earlier papers for a discussion of this model. Our likelihood analysis
shows that reionization is 50 (99) \% complete between redshifts
$z = $7.9 -- 11.6 (5.8 -- 7.8) at 95\% confidence level. Hence, under the
assumptions made in the model, we find that completion of reionization 
cannot occur earlier than $z \approx 8$, essentially ruling out models of 
very early
reionization. The reason for this is that the number of photons in the IGM
at $z = 6$ 
is very low as implied by the Ly$\alpha$ forest data. In order to take the data point into account, the models typically
cannot have too high a emissivity at $z \sim 6$.
On the other hand, the constraints on $\tau_{\rm el}$ imply that reionization
must be initiated early enough. Thus the IGM
has to go through a gradual reionization phase. As we discussed above, 
the gradual reionization
is maintained by a combined action of radiative and chemical feedback effects.

Interestingly, we find that a couple of data points for
$\de N_{\rm LL}/\de z$ lie above the 2-$\sigma$ limits of our analysis. 
In models where these points agree with the data, the photon mean free path
$\lambda_{\rm mfp}$, and hence the photoionization rate
$\Gamma_{\rm PI}$, are relatively smaller. These lead to larger GP optical
depths which then violate the Ly$\alpha$ forest constraints.
This discrepancy can arise either (i) because of 
some unaccounted systematics present
in the data or (ii) from the simplifying assumptions made in our models for
calculating $\lambda_{\rm mfp}$. The actual reason needs to be investigated
further.

\section{Principal Component Analysis}

\subsection{Motivation}

It is most likely that the star-forming efficiencies and escape fractions and hence $N_{\rm ion}$ are functions of halo mass and redshift; 
however since the dependencies are not well understood, they were
taken to be constant for each considered stellar population in the previous Section. 
The question one can ask is that how would the constraints on reionization
histories of the previous Section change when the evolution of
$N_{\rm ion}$ is taken into account.
Ideally one would like 
to do a rigorous likelihood analysis with $N_{\rm ion}$ varying with $z$ and 
see the possible ranges of 
reionization histories consistent with available data.
One possible approach could be to parameterize $N_{\rm ion}(z)$ using some
(known) function and constrain the parameters of the function
\cite{2010MNRAS.408...57P}. However, it is
possible that the reionization constraints thus obtained could depend 
on the nature of the function chosen. In addition, it is not clear as to
how many parameters should be used to parameterize the function. 

An alternative approach is to assume
$N_{\rm ion}(z)$ to be completely arbitrary and decompose it into principal
components. 
These principal components essentially filters out components
of the model which are most sensitive to the data. Obviously, these components
are the ones which can be
constrained most accurately, while the others cannot be done so. This 
principal component analysis (PCA), thus, should 
give an idea as to which aspects of $N_{\rm ion}$ can be constrained with
available data. This implies that one should get a clear idea about the 
optimum number of parameters required to model $N_{\rm ion}$ to fit the
data most accurately. 

In order to carry out such analysis, we modify the model described in the
previous Section in following respects:

\begin{itemize}

\item We take $N_{\rm ion}$ to be a function of $z$. Unlike in the 
previous Section, we do not explicitly assume the presence of two
population of stars but rather we include only one stellar population; any change
in the characteristics of these stars over time would be accounted for 
in the evolution of $N_{\rm ion}$.

\item Clearly, the chemical feedback prescription has to abandoned in this
model, as there are no two different populations of stars anymore. 
The chemical feedback is rather taken into account indirectly by the evolution of 
$N_{\rm ion}$. However, we retain radiative feedback in the model given its
weak dependence on the specific stellar population properties.

\end{itemize}

In recent years there has been a wide use of this method in cosmological data analysis. 
The first set of works were mostly related to CMB data where, e.g., 
\citeN{1999MNRAS.304...75E} and 
\citeN{2002MNRAS.332..193E} used principal component analysis of CMB anisotropy measurements to investigate degeneracies among 
cosmological parameters.
\citeN{2005PhRvD..72b3510K} applied PCA to study how accurately CMB observables can constrain inflaton potential in a model-independent manner. 
\citeN{2006MNRAS.372..646L} used PCA techniques for measuring departures from scale-invariance in the primordial 
power spectrum of density perturbations using cosmic microwave background (CMB) $C_l$ data. 
\citeN{2008ApJ...672..737M} developed a model-independent method to study the effects of reionization on the large-scale E-mode polarization for 
any reionization history with the help of principal component analysis followed by the earlier work by \citeN{2003PhRvD..68b3001H}.
In the context of weak lensing surveys, \citeN{2006A&A...452...63M}
studied the degeneracies between cosmological parameters and measurement errors from cosmic shear surveys using PCA. 
The PCA has also been employed as an effective tool 
in the context 
of type Ia supernova observations to constrain the equation of state of dark energy \cite{2003PhRvL..90c1301H,2005PhRvD..71b3506H,2009JCAP...12..025C,2010PhRvL.104u1301C}.

\subsection{Basic theory of PCA}

Consider a set of $n_{\rm obs}$ observational data points 
labeled by ${\cal G}_{\alpha},~\alpha=1,2,\ldots,n_{\rm obs}$. 
Recall that ${\cal G}_{\alpha}$ can represent combinations of different data sets, e.g., in our case ${\cal G}_{\alpha} = \{\log(\Gamma_{\rm PI}), \tau_{\rm el}, \de N_{\rm LL}/\de z\}$.

Now, let us assume that our model contains an unknown function $N_{\rm ion}(z)$,
which we wish to constrain through observations. We can divide our entire 
redshift interval $[z_{\rm min}, z_{\rm max}]$ into (equal) bins of width
$\Delta z$ and represent $N_{\rm ion}(z)$ by a set of $n_{\rm bin}$ discrete
free parameters 
\begin{equation}
 N_{\rm ion}(z_i)\equiv N_i;~~i=1,2,...,n_{\rm bin}
\end{equation}
where
\begin{equation}
 z_i=z_{\rm min}+(i-1)\Delta z
\end{equation}
and the bin width is given by
\begin{equation}
 \Delta z=\frac{z_{\rm max}-z_{\rm min}}{n_{\rm bin}-1}.
\end{equation}
In other words, we have modelled reionization using the value of 
$N_{\rm ion}$ in each redshift bin. 
We can also include other free parameters apart from 
$N_{\rm ion}(z_i)$ in the analysis, like the normalization of the mean free path $\lambda_0$, cosmological parameters etc. However, for the moment let us 
assume that these parameters are fixed (known from other observations) and
concentrate on $N_{\rm ion}(z_i)$ only. We will address the inclusion of other
parameters later in this Section.

The next step is to assume a fiducial model for $N_{\rm ion}(z_i)$, which we
denote by $N_{\rm ion}^{\rm fid}(z_i)$. The fiducial model should be chosen such
that it is close to the ``true'' model. The departure from the 
fiducial model is denoted by
\begin{equation}
 \delta N_{\rm ion}(z_i)=N_{\rm ion}(z_i)-N_{\rm ion}^{\rm fid}(z_i)\equiv \delta N_i.
\end{equation}
We can then construct the $n_{\rm bin}\times n_{\rm bin}$ Fisher matrix  
\begin{equation}
 F_{ij}=\sum_{\alpha=1}^{n_{\rm obs}}\frac{1}{\sigma_{\alpha}^2}\frac{\partial {\cal G}_{\alpha}^{\rm th}}{\partial N_i}\frac{\partial {\cal G}_{\alpha}^{\rm th}}{\partial N_j},
\label{eq:fisher_matrix}
\end{equation}
where ${\cal G}_{\alpha}^{\rm th}$ is theoretical value of 
${\cal G}_{\alpha}$ modelled using the $N_i$ and $\sigma_{\alpha}$ is the
observational error on ${\cal G}_{\alpha}$. The derivatives in the 
above relation are evaluated at the
fiducial model $N_i = N^{\rm fid}_i$.\footnote{It is worthwhile to mention that any analysis based on the  Fisher matrix $F_{ij}$, in principle, depends on the fiducial model 
chosen. The principal component analysis, which essentially involves 
diagonalizing $F_{ij}$, is thus dependent on the choice of $N^{\rm fid}_i$ too. In this sense, the PCA is not completely 
model-independent.}

Once the Fisher matrix is constructed, we can determine its eigenvalues and corresponding eigenvectors. The principal value decomposition is 
then given by the eigenvalue equation 
\begin{equation}
 \sum_{j=1}^{n_{\rm bin}} F_{ij}S_{jk}=\lambda_k S_{ik}
\end{equation}
where $\lambda_k$ are the eigenvalues and the eigenfunctions corresponding to $\lambda_k$ are the $k$-th column of the matrix $S_{ik}$, 
these are the principal components of $N_i$. They can be thought of a 
function of $z$ i.e., $S_{ik}=S_{k}(z_i)$.

The eigenvalues $\lambda_k$ are usually ordered such that
$\lambda_1 \geq \lambda_2 \geq \ldots \geq \lambda_{n_{\rm bin}}$,
i.e., $\lambda_1$ corresponds to the largest eigenvalue while 
$\lambda_{\rm nbin}$ the smallest.
The eigenfunctions are both orthonormal and complete 
and hence we can expand any function of $z$ as linear combinations of them. 
In particular we can expand 
the departure from the fiducial model as
\begin{equation}
 \delta N_i=\sum_{k=1}^{n_{\rm bin}} m_kS_{k}(z_i); ~~
m_k=\sum_{i=1}^{n_{\rm bin}} \delta N_{\rm ion}(z_i)S_k(z_i)
\end{equation}
where $m_k$ are the expansion coefficients with $m_k = 0$ for 
the fiducial model. We can now describe our model by the
coefficients $m_k$ rather than the original parameters $\delta N_i$.
The advantage is that, unlike $N_i$, the coefficients $m_k$ are uncorrelated 
with variances given 
by the inverse eigenvalue:
\be
\langle m_i~ m_j \rangle
= \f{1}{\lambda_i} \delta_{ij}
\e
 The accuracy with which we can determine $\delta N_{\rm ion}$ at a particular $z_i$ is determined by the Cramer-Rao bound 
\begin{equation}
 \left\langle\delta N^2_{\rm ion}(z_{i})\right\rangle\geq\sum_{k=1}^{n_{\rm bin}}\frac{S_{k}^2(z_{i})}{\lambda_{k}}
\end{equation}
So, the largest eigenvalues correspond to minimum variance. The eigenvalues which are smaller would essentially increase the uncertainty in determining 
$\delta N_{\rm ion}(z_i)$. 
Hence, most of the information relevant for the observed data points
${\cal G}_{\alpha}$ is contained in the first few modes with the largest
eigenvalues. One may then attempt
to reconstruct the function $\delta N_{\rm ion}(z_i)$ using only the first 
$M \leq n_{\rm bin}$ modes:
\begin{equation}
  \delta N_{i}^{(M)}=\sum_{k=1}^{M} m_{k}S_{k}(z_{i}).
\end{equation}
However, in neglecting the last $n_{\rm bin} -M$ terms, one introduces a bias
in determining $\delta N_{\rm ion}(z_i)$. One has to then use a carefully
chosen $M$ to perform the analysis; the choice usually depends on the
particular problem in hand. We shall discuss our choice of $M$ in the next
Section.

In realistic situations, there will be other free parameters (apart from
$m_k$ or $\delta N_i$) in the model; these could be, e.g., 
the normalization of the mean free path $\lambda_0$, cosmological
parameters etc. Let there be $n_{\rm ext}$ number of extra parameters 
other than $m_k$; this means that we are now dealing with
a total of $n_{\rm tot} = n_{\rm bin} + n_{\rm ext}$ parameters.
In this case, we can still form the Fisher matrix of 
$n_{\rm tot} \times n_{\rm tot}$ dimensions which can be written as
\be
{\cal F} = \left(\begin{array}{cc}
{\matrixsymbol F} & {\matrixsymbol B}\\
{\matrixsymbol B}^T & {\matrixsymbol F'}
\end{array}\right)
\e
where ${\matrixsymbol F}$ is the $n_{\rm bin} \times n_{\rm bin}$-dimensional 
Fisher matrix for the $\delta N_i$, ${\matrixsymbol F'}$ is the
$n_{\rm ext} \times n_{\rm ext}$-dimensional 
Fisher matrix for the other parameters and 
${\matrixsymbol B}$ is a $n_{\rm bin} \times n_{\rm ext}$-dimensional
matrix containing the cross-terms.
One can then invert the above ${\cal F}$ to obtain the corresponding Hessian
matrix ${\cal T} = {\cal F}^{-1}$. Following that, one simply retains 
the sub-block ${\matrixsymbol T}$ corresponding to $\delta N_i$
whose principal components will be ``orthogonalized'' to
the effect of the other parameters. The resulting ``degraded'' sub-block
will be \cite{1992nrfa.book.....P}
\be
{\matrixsymbol \tilde{F}} = {\matrixsymbol T}^{-1}
= {\matrixsymbol F} - {\matrixsymbol B} {\matrixsymbol F'}^{-1} {\matrixsymbol B}^T
\e

In this work we keep the cosmological parameters fixed; however we still
need to use the 
above formalism to marginalize over $\lambda_0$. In that
case, obviously $n_{\rm ext}=1$.

\section{Results}

The detailed results of our PCA are presented in this Section. 

\subsection{Fiducial model}

The first task is to make an assumption for the fiducial model
 $N_{\rm ion}^{\rm fid}(z)$. The model
should match the $\Gamma_{\rm PI}$ and $\de N_{\rm LL}/\de z$ data points
at $z < 6$ and also produce a $\tau_{\rm el}$ in the acceptable range.
Unfortunately, the simplest model with $N_{\rm ion}$ being constant does not
have these requirements (recall models with only PopII stars were disfavoured
in the previous Section). We have found earlier that 
the effective $N_{\rm ion}$ should be higher at early epochs dominated
by PopIII stars and should approach a lower value at $z \sim 6$ determined
by PopII stars. In this work we take $N_{\rm ion}^{\rm fid}$ 
to be the model given by mean values of the free parameters
in Section \ref{sec:reion_constraints}.

\begin{figure}
\includegraphics[width=0.45\textwidth,angle=0]{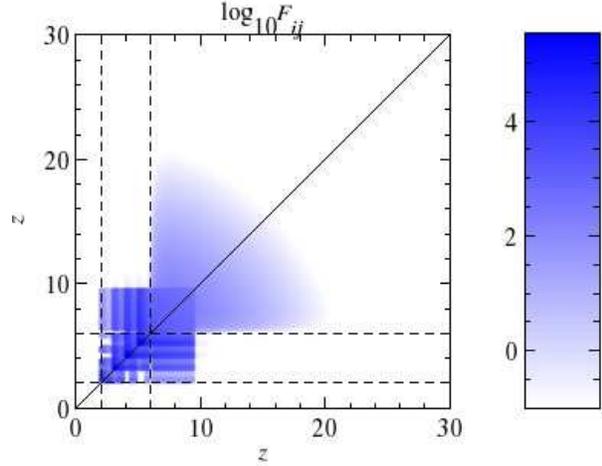}
  \caption{The Fisher matrix $F_{ij}$ in the $z-z$ plane.}
\label{fig:fisherplot}
\end{figure}

The choice of this $N^{\rm fid}_{\rm ion}$ may seem somewhat
arbitrary as there could be many other forms of $N_{\rm ion}$ which may
match the data equally well.
We have chosen this to be our fiducial model because of the following reasons:
(i) it is obtained from a physically-motivated model of star formation
which includes both metal-free and normal stars,
(ii) it is characterized by a higher $N_{\rm ion}$ at higher redshifts and hence produces a good match with different observations considered in this
work,  and (iii) the transition
from higher to lower values is smooth (i.e., there is no abrupt transition
or sharp features). The final conclusions of this work (to be presented later
in the Section) would hold true for any fiducial
model having these three properties (though the actual functional form might
be different). The match with the data for our fiducial model is similar
to Fig. 2 of \citeN{2009arXiv0904.4596C}.

We have run the reionization models over a redshift range
$[z_{\rm min}:z_{\rm max}] = [0:30]$, with a bin width of 
$\Delta z = 0.2$. This gives $n_{\rm bin} = 151$. We have checked and found
that our main conclusions are unchanged if we vary the bin width between
0.1--0.5.

The Fisher matrix $F_{ij}$ defined in equation (\ref{eq:fisher_matrix}) is evaluated 
at the fiducial model and is shown as
a shaded plot in the $z-z$ plane in Figure \ref{fig:fisherplot}. 
Firstly, the components of the the matrix vanish for 
$z < 2$ because there are no data points considered at these redshifts.
The plot shows different characteristics for $F_{ij}$ at redshift intervals 
$2 < z < 6$ and $z > 6$. For $z < 6$, the values of $F_{ij}$ are considerably
higher because it is determined by the sensitivity of $\Gamma_{\rm PI}$ and 
$\de N_{\rm LL}/\de z$ on $N_{\rm ion}$ 
 and it turns out that $\Gamma_{\rm PI}$ is extremely
sensitive to changes in $N_{\rm ion}$. One can see a band-like structure in the 
information matrix which essentially corresponds to the presence 
of data points. The
regions where data points are sparse (or non-existent, like between 
$z=2$ and 3), the
value of $F_{ij}$ is relatively smaller, implying that one cannot
constrain $N_{\rm ion}$ from the data in those redshift bins.
On the other hand, the information at $z > 6$ is 
determined by the sensitivity of $\tau_{\rm el}$ on 
$N_{\rm ion}$. Once can see that $F_{ij} \to 0$ at the highest redshifts
considered; this is expected because the collapsed fraction of haloes
is negligible at those redshifts and hence there exist no free electrons to
contribute to $\tau_{\rm el}$. The precise redshift range at which 
$F_{ij}$ become negligible depends on the (measured) value of 
$\tau_{\rm el}$. For the WMAP7 measurements, we find that 
$F_{ij}$ is negligible for $z > 14$; if, e.g., the measured value of 
$\tau_{\rm el}$ were higher, 
$F_{ij}$ would be non-negligible till relatively higher redshifts. We can
thus conclude that it is not possible to
constrain any parameters related to star formation at redshifts $z > 14$
using the data sets we have considered in this work.

\begin{figure}
  \includegraphics[height=0.45\textwidth, angle=270]{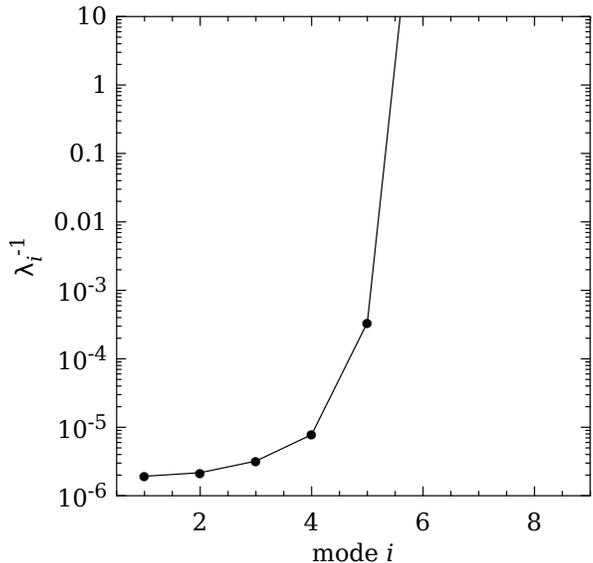}
  \caption{The inverse of eigenvalues which essentially measures the variance on
the corresponding coefficient $m_i$. For modes larger than 5, the eigenvalues
are extremely small.}
\label{fig:eigenval-plot}
\end{figure}

Once we diagonalize the matrix $F_{ij}$, we obtain its eigenvalues and the
corresponding eigenmodes. The inverse of the eigenvalues, which are
essentially the variances of the corresponding modes, 
are plotted in Figure \ref{fig:eigenval-plot}.
Since the eigenvalues $\lambda_i$ are sorted in 
ascending order, the variances are larger for higher modes.
For modes $i > 5$, the eigenvalues are almost zero and the variances
are extremely large. This implies that the errors on $N_{\rm ion}$ would increase
dramatically if we include modes $i > 5$. 

\begin{figure}
  \includegraphics[height=0.45\textwidth, angle=-90]{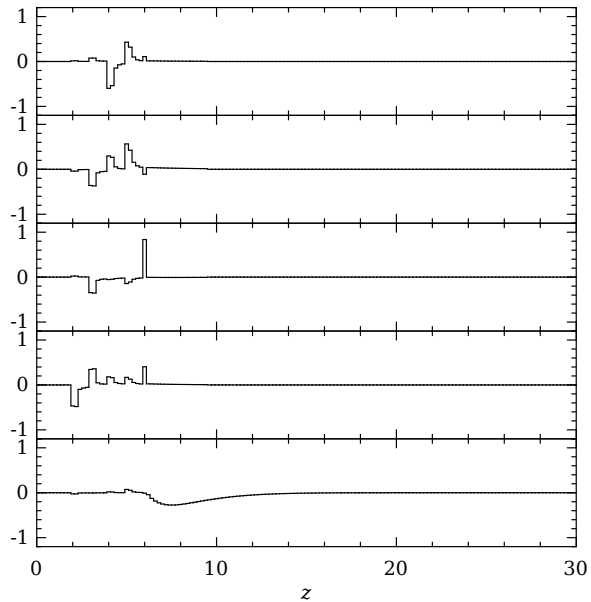}
  \caption{The first 5 eigenmodes of the Fisher matrix, i.e., $S_k(z);~k=1,\ldots,5$.}
\label{fig:eigenvec-plot-1}
\end{figure}

The first 5 eigenmodes which have the lowest variances are shown in Figure
\ref{fig:eigenvec-plot-1}. Clearly, all these modes
tend to vanish at $z > 14$, which is because of
$F_{ij}$ being negligible at these redshifts. Also, modes are identically zero at
$z < 2$ because we have not used any data points at these redshifts. 
The first 4 modes 
essentially trace the sensitivity of $\Gamma_{\rm PI}$ and $\de N_{\rm LL}/\de z$ at $z < 6$ on the value of
$N_{\rm ion}$.
One can see a number of spikes and troughs in these modes whose positions correspond
to the presence of data points and amplitudes correspond to the error-bars 
on these data points (smaller the error, larger the amplitude). 
The shape of the 5th mode is vary much different from the previous four.
This mode essentially contains 
the behavior of $N_{\rm ion}$ at $z > 6$
and hence it characterizes the
sensitivity of $\tau_{\rm el}$ on $N_{\rm ion}$. 
Since $\tau_{\rm el}$ is obtained by integrating the reionization history 
over the whole redshift range, the sensitivity covers a wide range
of redshifts (which is unlike the sensitivity of $\Gamma_{\rm PI}$).
The sensitivity is maximum around $z \approx 8$, 
which is determined by the nature of the fiducial model. The sensitivity
falls at $z > 8$ because there is a reduction in the number of
sources and free electrons. Interestingly the sensitivity falls at 
$z < 8$ too which is due to the fact that reionization
is mostly complete at these redshifts $x_e \to 1$ 
and hence changing $N_{\rm ion}$ does not change the value of $\tau_{\rm el}$
significantly.

The modes with smaller eigenvalues have large variances and hence introduce huge uncertainties in the determination of $N_{\rm ion}$. The
modes are characterized by sharp features at different redshifts and they
do not contain any significant information about the overall reionization
history.

\subsection{Choice of the number of modes}
\label{sec:choice_of_M}

\begin{figure*}
\includegraphics[height=\textwidth, angle=270]{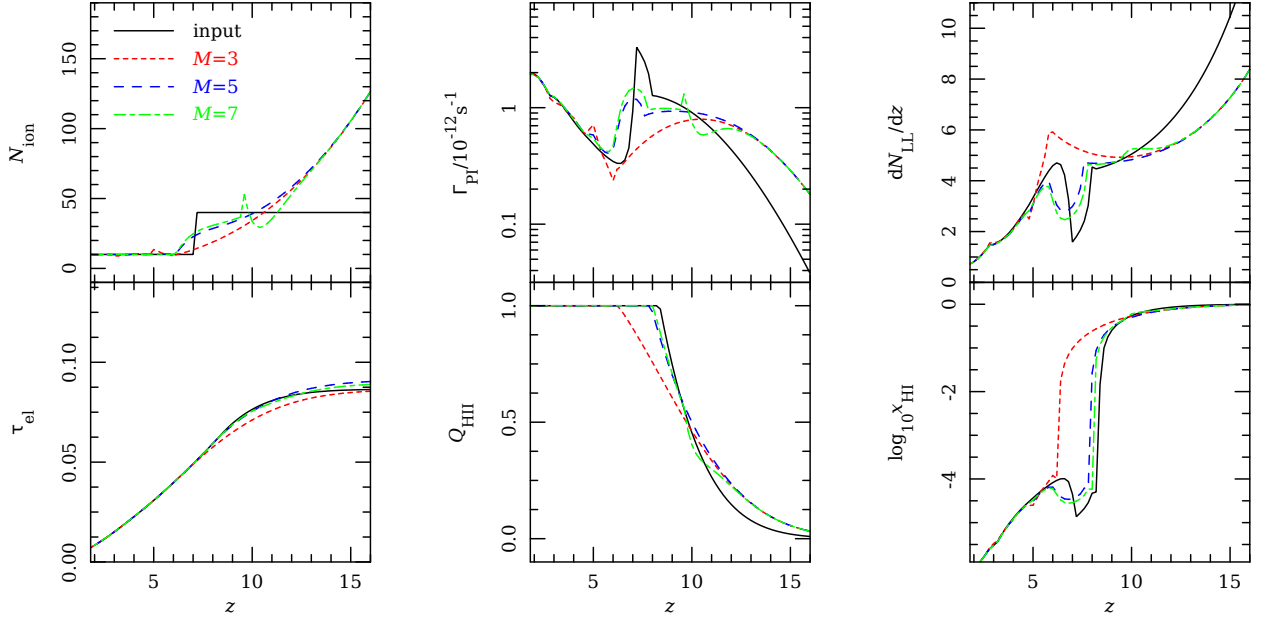}
\caption{Recovery of various quantities related to reionization when the input 
underlying model of $N_{\rm ion}(z)$ is assumed to be a step function (shown by
solid lines). The extent of recovery is shown when the first 
3 (short-dashed lines), 5 (long-dashed lines), 7 (short-long-dashed lines) 
PCA modes are included in the analysis.}
\label{fig:reconstruct}
\end{figure*}

The next step in our analysis is to decide on how many modes $M$ to use.
In the case where
$M = n_{\rm bin}$, all the eigenmodes are included in the analysis and 
no information is thrown away. However, this would mean that 
modes with very small eigenvalues (and hence large uncertainties) 
are included and thus the errors in recovered quantities would be
large. Reducing $M$ is accompanied
by a reduction in the error, but an increased chance of
getting the recovered quantities wrong (which is known as bias).

It is thus natural to ask
what could be the optimum value of $M$ for calculations. The
most straightforward way, which is used often, is to 
determine it by trial and error, i.e., more and more terms are added
till one gets some kind of convergence in the recovered quantities 
\cite{2008ApJ...672..737M}. 
Let us first work out the simplistic trial-and-error approach
to fix $M$ and as we shall see that this would be 
helpful in understanding recovery of various parameters using
PCA.
We have already discussed that  inclusion of modes $> 5$ implies 
drastic rise in the errors. Hence,
it seems that $M \leq 5$ would be a good choice. The question is
whether throwing away such a large number of modes ($n_{\rm bin} -M$) 
would introduce large biases in the recovered quantities.

In order to examine these issues in more detail, let us assume that
the underlying ``true'' form of $N_{\rm ion}$ is very different from 
the fiducial model
we have chosen and then try to estimate the errors we make in recovering this
underlying model using only the first few modes. 
In order to put our method to test, it
is then natural to assume an underlying model which is
noticeably different from the fiducial one and study its recovery
using only the first few modes. 
Recall that the fiducial model represents a smoothly varying
$N_{\rm ion}$, so we assume the underlying input model to be 
one having an abrupt transition, e.g., 
a step function:
\be
\begin{array}{lcll}
N_{\rm ion}^{\rm inp}(z) &=& 10 & \mbox{ for } z < 7 \nline
&=& 40 &  \mbox{ for } z \geq 7
\end{array}
\label{eq:step}
\e
The parameters in the model are adjusted so that it matches with 
observations of $\Gamma_{\rm PI}$ and $\de N_{\rm LL}/\de z$ at $z < 6$ and
also gives the correct observed value of $\tau_{\rm el}$. The idea
would be to check whether we are able to recover quantities of
interest with reasonable accuracy with $M=5$. The model chosen 
above is similar to the abrupt-transition model considered in
\citeN{2005MNRAS.361..577C}.

\begin{table}
\begin{tabular}{c|c|c|c|c}
Parameters & Input value & \multicolumn{3}{|c|}{Recovered values}\\
&&$M=3$&$M=5$&$M=7$\\
\hline
$z(Q_{\rm HII}=0.5)$ & $9.817$ & $9.722$ & $10.004$ & $9.728$\\
$z(Q_{\rm HII}=0.99)$ & $8.337$ & $6.311$ & $7.875$ & $8.037$\\
\hline
\end{tabular}
\caption{The recovered quantities for a input model where $N_{\rm ion}$ is
represented by a step function when only the first $M$ PCA modes are included
in the analysis.
}
\label{tab:reconstruct}
\end{table}

The results of our analysis are shown in Figure \ref{fig:reconstruct}
and in Table \ref{tab:reconstruct}. 
In the figure, we have plotted, as functions of redshifts, the four 
quantities relevant to reionization which we would like to recover, 
namely, $N_{\rm ion}$ (top-left panel), 
the photoionization rate $\Gamma_{\rm PI}$ (top-right panel), 
the volume filling factor of ionized regions $Q_{\rm HII}$ (bottom-left panel) 
and the globally averaged neutral hydrogen fraction 
$x_{\rm HI}$ (bottom-right panel). Different curves represent the input
step model (solid) and the recovered quantities for 
three values of $M = 3,5,7$ (short-dashed, long-dashed, short-long-dashed, respectively). 
We have not shown results for intermediate values of $M$ (i.e., $M = 4,6$) 
because the difference between successive plots is too small to be noticed.
It is clear from the top left panel that the recovered
$N_{\rm ion}$ is excellent for $z < 6$ because the fiducial and input models 
agree at these redshifts, which is a manifestation of the fact that 
the value of $N_{\rm ion}$ is highly constrained by good quality data points
at these redshifts. On the other hand, the recovery is quite
poor for $z > 6$. This is because the
evolution at $z > 6$ is only weakly constrained by $\tau_{\rm el}$. 
In particular at $z > 14$, the modes are essentially zero and hence all
models tend to the fiducial one implying that it is impossible to recover
$N_{\rm ion}$ at $z > 14$ with the first few modes.  

The top-middle and top-right panels show the
corresponding plots for the photoionization rate $\Gamma_{\rm PI}$ and the
redshift distribution of Lyman-limit systems $\de N_{\rm LL}/\de z$ respectively. 
The input model has a sharp feature
around $z \approx 7$ in both the quantities 
arising mainly from the abrupt step in $N_{\rm ion}$. 
The reionization is complete ($Q_{\rm HII} = 1$) at $z \approx 8$ after 
which the photoionization
rate rises sharply because of overlap of ionized regions and consequent 
rise in mean free path (which manifests itself as a sharp drop in the
number of LLS). This rise in $\Gamma_{\rm PI}$ is suddenly halted at $z = 7$ 
where we see
a sharp decline because of the corresponding step decline in $N_{\rm ion}$. 
Following that, $\Gamma_{\rm PI}$ settles to a smaller value 
(corresponding to a smaller value of $N_{\rm ion}$) and subsequently 
shows a gradual rise arising again from the rise in mean free path. 
Interestingly, this feature is completely missing in the 
recovered model for $M = 3$ (and also for $M=4$, not shown in the figure). 
The feature shows up
when $M$ is increased to 5, though the exact nature of this 
feature is not identical to the input one. Increasing $M$ to 7
introduces other sharp features at $z \sim 10$ which are not present in the 
input model. Of course, the recovery at $z > 14$ is poor as
most of the eigenmodes hardly contain any information at these redshifts and the
recovered models simply follow the fiducial model. 
Hence, the recovery of the photoionization rate and the LLS distribution 
is probably not satisfactory overall,
however we can recover it
with reasonable accuracy for $z < 12$ by considering the first $M = 5$ modes.

The recovery of $\tau_{\rm el}$ is shown in the bottom-left panel. It is clear
that the recovery is good for all values of $M$.
In most reionization studies, the quantities of main interest are
the $Q_{\rm HII}$ and $x_{\rm HI}$, which are plotted in the bottom-middle and
bottom-right panels respectively. One can easily see from both the panels 
that that the agreement
between the $M=3$ case and the input model is quite poor (which is the case
for $M=4$ as well). 
In particular, reionization is complete at $z \approx 8.5$ for the input model,
while it completes only at $z \approx 6$ for $M=3$ case 
(see Table \ref{tab:reconstruct}).
However, the
moment $M$ is increased to 5, one has a remarkable match with the
input model, e.g., the difference in $Q_{\rm HII}$ is $< 0.1$ for 
$z<12$ while the difference is $<10\%$ for $z < 10$. 
Unfortunately, we cannot recover the sharp feature in $x_{\rm HI}$ around $z = 7$ 
for the input model
(which corresponds to a similar feature in $\Gamma_{\rm PI}$, discussed above)
for the $M=5$ case, however the overall agreement with the input model
is still quite good.
The agreement is further improved as
we increase $M$ (to 7 in the plot) but that comes at the cost of increasing errors. As far as
recovering the basic reionization history (i.e., evolution of 
$Q_{\rm HII}$ and $x_{\rm HI}$) is concerned, $M=5$ seems to be the
optimum choice. 

\begin{figure}
\includegraphics[height=0.45\textwidth, angle=270]{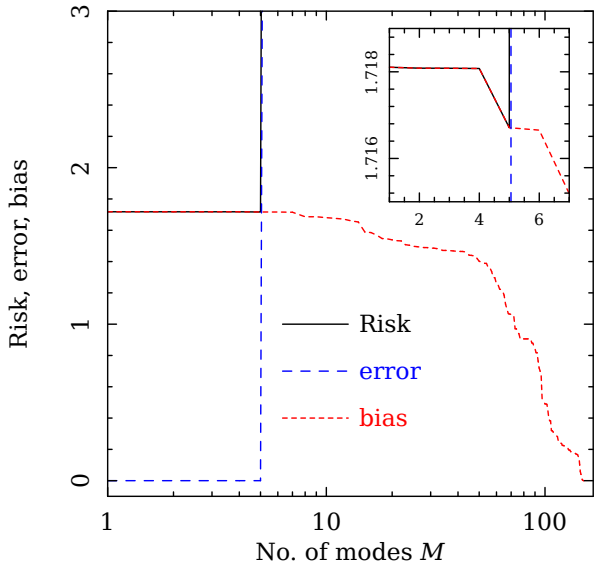}
\caption{Dependence of Risk, error and bias as defined in equation (\ref{eq:risk}) on the number of modes $M$. The blow-up of a region around $M = 5$ is shown 
in the inset which shows that there is a clear minimum in the Risk at $M=5$.}
\label{fig:risk}
\end{figure}

It is important to point out that the recovery of various quantities
related to reionization is good (or excellent, in some cases) even when
the recovered value of $N_{\rm ion}$ is incorrect. 
This may seems surprising as it is the value of $N_{\rm ion}$ that acts as
a source for reionization.
To understand this apparent paradox, note
that the recovery of $N_{\rm ion}$ is poor mostly at $z > 12$. At these
redshifts the collapsed fraction $\de f_{\rm coll}/\de t$ is typically
small, hence the source emissivity
$N_{\rm ion} \de f_{\rm coll}/\de t \to 0$ at these epochs. Hence even if we
change the value of $N_{\rm ion}$, the absolute change in the emissivity is
negligible and hence the reionization process remains relatively unaffected.
There is another way of looking at it: the extent of recovery of various
quantities at $z > 6$ is determined by the 
behaviour of PCA modes at $z > 6$ which, in turn, is determined by 
the data set related to $\tau_{\rm el}$. Now $\tau_{\rm el}$ is most
sensitive to the ionized fraction
$Q_{\rm HII}(z)$ at $z > 6$. Hence, it is not surprising that 
$Q_{\rm HII}(z)$ would be nicely recovered at these redshifts. Such arguments
can be extended for other quantities too. This also brings out the fact 
that in order to recover $N_{\rm ion}(z)$ (and thus star formation, escape fraction and chemical feedback) reliably, one requires data points at 
$z > 12$ related to quantities which are 
sensitive to $N_{\rm ion}$, like say, hypothetically, a good constraint
on $\Gamma_{\rm PI}$ at $z \sim 12$ can constrain $N_{\rm ion}$ at those redshifts.

To summarize our results on recovering the input step model, 
the recovery of all the quantities is excellent for $z < 6$.
We find that
recovery of $N_{\rm ion}$ at $z > 6$ is not satisfactory. 
The recovery of $\Gamma_{\rm PI}$ at $z < 12$ is quite reasonable 
by considering the first $M=5$ modes. 
Fortunately, the recovery of $Q_{\rm HII}$ and $x_{\rm HI}$ 
turns out to be excellent 
for $M = 5$. 
Hence we can use the coefficients $m_i$ of these 5 best constrained 
eigenmodes as our model parameters instead of $N_{\rm ion}(z_i)$ without 
significant loss of 
information.

We should mention that the above analysis depends on the 
choice of the input model
which is taken to be the step function. In fact, 
the recovery is better if the input $N_{\rm ion}$ is a smoother function
(provided it satisfies the observational constraints, of course). In 
particular, all models which are bracketed by the fiducial model and 
the step model would end up giving good agreements for 
$\Gamma_{\rm PI},\de N_{\rm LL}/\de z, \tau_{\rm el}, Q_{\rm HII}$ and $x_{\rm HI}$. 
Of course, if the input models
have sharp features at some particular redshift(s) $z > 6$, 
those features may not be
recovered satisfactorily by including only first few terms.

A slightly more formal approach is to estimate $M$ 
by minimizing the quantity Risk, which is defined as \cite{2001astro.ph.12050W} 
\begin{equation}
  R=\sum_{i=1}^{n_{\rm bin}}\left(\delta N_{i}^{(M)}\right)^{2}+\sum_{i=1}^{n_{\rm bin}}\left\langle\left(\delta N_{i}^{(M)}\right)^{2}\right\rangle
\label{eq:risk}
\end{equation}
The 1st term in the RHS is the bias contribution which arises from neglecting
the higher order terms, and the 2nd term is the uncertainty given 
by Cramer-Rao bound  which rises as higher order terms (i.e., those corresponding to smaller eigenvalues) are included:
\begin{equation}
  \left\langle\left(\delta N_{i}^{(M)}\right)^{2}\right\rangle \geq \sum_{k=1}^{M}\frac{S_{k}^2(z_{i})}{\lambda_{k}}
\end{equation}
However, the calculation of Risk, as defined above, involves assumption of an 
``underlying model'', hence the determination of $M$ using this method would
be model-dependent. Let us assume the underlying model to be the same as
equation (\ref{eq:step}). Then the dependence of the Risk on the number
of modes $M$ is shown in Figure \ref{fig:risk}. In addition, we also
show the plots of bias [first term of the rhs in equation (\ref{eq:risk})]
and the error [second term of the rhs in equation (\ref{eq:risk})] are also
shown. It is clear that the value of error is small for lower $M$ which 
is a direct consequence of small eigenvalues. The error shoots up drastically
for $M > 5$ which is what we discussed in the previous Section.
On the other hand, the bias is higher for small $M$ and decreases
gradually as more and more terms in the summation are included. The Risk, which
is the sum of these two quantities, has a clear minimum at $M = 5$ (which
is more clear from the inset in Figure \ref{fig:risk}). Hence we conclude
that $M=5$ is the optimum value to be used.

The main conclusion of this Section is that one needs five parameters
to describe the reionization history which can be constrained with 
the data considered in this paper. Out of these five, four parameters 
are required to describe the emissivity at $z < 6$ where most of the
data points exist; these parameters are the best-constrained ones. 
The fifth parameter characterizes the evolution of $N_{\rm ion}$ at $z > 6$
and is essentially determined by the WMAP constraints of $\tau_{\rm el}$. 
Inclusion of more parameters would lead to overfitting of the data and
hence the constraints on the parameters would be highly uncertain.

\subsection{Constraints on reionization history}

\begin{figure*}
\includegraphics[height=\textwidth, angle=270]{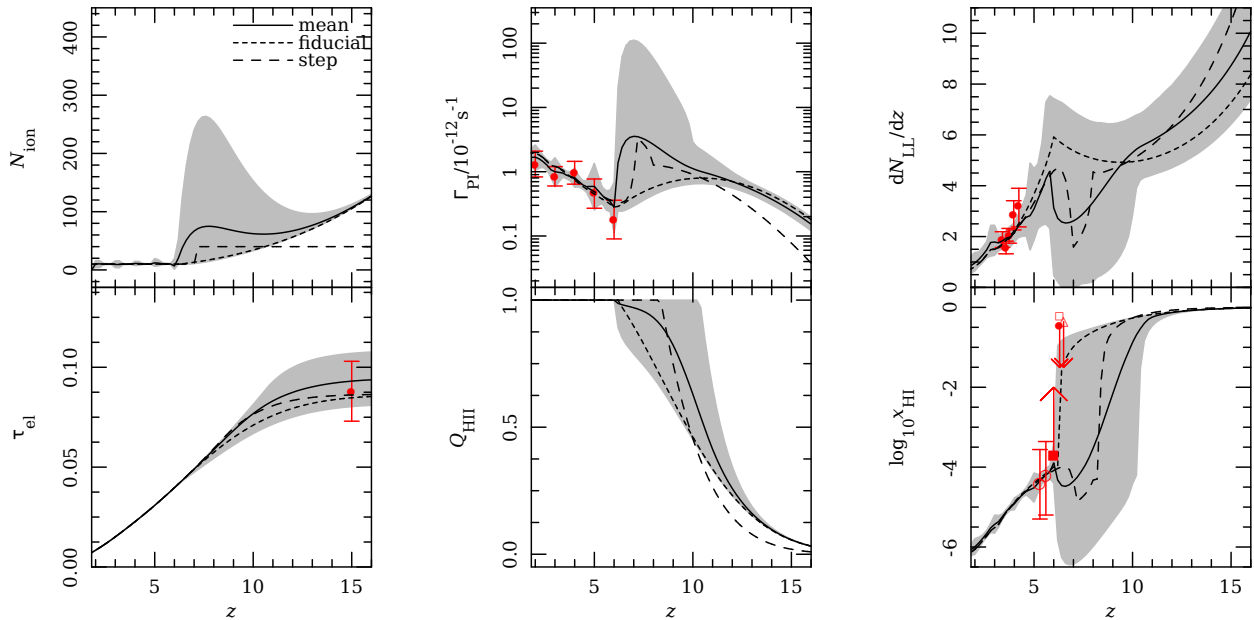}
\caption{The marginalized posteriori distribution of various 
quantities related to
reionization history obtained from the PCA. The different quantities in the
panels are identical
to those in Figure \ref{fig:threep_posteriori}.
The solid lines correspond to the model described by mean values of the
parameters
while the shaded regions correspond to 2-$\sigma$ limits. 
In addition, we show the properties of the 
fiducial model (short-dashed lines) and the 
step model used in Section \ref{sec:choice_of_M} (long-dashed lines).
The points with error-bars denote the observational data points
which are identical to those in Figure \ref{fig:threep_posteriori}.
}
\label{fig:PCA_posteriori}
\end{figure*}

\begin{table}
\begin{tabular}{c|c|c}
Parameters & Mean value & 95\% confidence limits\\
\hline
$m_1$ & $0.002$ & $[-0.002, 0.018]$\\
$m_2$ & $0.007$ & $[-0.001, 0.032]$\\
$m_3$ & $-0.003$ & $[-0.012, 0.004]$\\
$m_4$ & $0.003$ & $[-0.003, 0.015]$\\
$m_5$ & $-0.065$ & $[-0.276, 0.003]$\\
$\lambda_0$ & $4.450$ & $[3.245, 5.906]$\\
\hline
$z(Q_{\rm HII}=0.5)$ & $10.349$ & $[9.528, 11.585]$\\
$z(Q_{\rm HII}=0.99)$ & $8.357$ & $[5.800, 10.270]$\\
\hline
\end{tabular}
\caption{The marginalized posterior probabilities
 with 95\% C.L. errors of all free parameters (top six parameters) and derived
parameters (from the seventh parameter down) for the reionization model
with principal component analysis.
}
\label{tab:PCA_likelihood}
\end{table}

The constraints on reionization are obtained by performing a 
Monte-Carlo Markov Chain (MCMC) analysis over the
parameter space of PCA amplitudes $\{m_1,\ldots,m_5\}$ and $\lambda_0$. The
cosmological parameters are kept fixed to the WMAP7 best-fit values. 
In order to carry out the analysis, we have developed a code based on
the publicly available  
COSMOMC\footnote{http://cosmologist.info/cosmomc/}
\cite{2002PhRvD..66j3511L} 
(which is widely used for running MCMC on CMB and other 
cosmological data sets).
To get accurate results from MCMC, we 
ensure that the parameter chains contain enough independent
samples over a sufficiently large volume of parameter space so that the
density of the samples converges to the actual posterior probability
distribution. We run a number of 
separate chains (varying between 5 to 10) until the Gelman and 
Rubin convergence statistics,
$R$, corresponding to the ratio of the variance of parameters between
chains to the variance within each chain, satisfies $R - 1 <
0.01$.

The mean values and the 95\% confidence limits on our parameters obtained
from our analysis are shown in Table \ref{tab:PCA_likelihood}. 
Our fiducial model $m_1=m_2=m_3=m_4=m_5=0$ is included within
the 95\% confidence limits of the parameters corresponding to the 
eigenmode amplitudes, however the mean values 
show clear departures from the fiducial model. This implies that the 
model characterized by the mean values of parameters, loosely mentioned 
as the ``mean model'' hereafter, is different from the 
fiducial one.

In order to see how different it is, we show the 
evolution of various quantities related to reionization is shown 
in Figure \ref{fig:PCA_posteriori}. The solid lines represent the mean model
while the shaded region correspond to 95\% confidence limits. For
comparison, we have also plotted the fiducial model (short-dashed) and 
the step model (long-dashed) which was introduced in Section \ref{sec:choice_of_M}.
We find that the fiducial model 
is within the 95\% confidence limits for the whole redshift range, while the
step model is within the 95\% confidence limits for $z < 10$.
Also note that the fiducial model
is actually near the edge of the shaded region, implying that there is 
a wide range of models allowed by the data which are 
characteristically different
from the fiducial model.

The next point to note is that all the quantities are highly constrained at
$z < 6$, which is expected as most of the observational 
information related to reionization
exists only at those redshifts. The errors also decrease at $z > 12$ as
there is practically no information in the PCA modes and hence all
models converge towards the fiducial one. 
This implies that early stages of reionization are almost similar
independent of the $N_{\rm ion}$ chosen.
The most interesting 
information regarding reionization is concentrated within a
redshift range $6 < z < 12$. 

It is very clear from the plot of $N_{\rm ion}(z)$ (top-left panel) that 
such quantity must necessarily increase from its constant value at $z < 6$. 
This rules out the possibility of reionization with a single stellar population 
having non-evolving IMF and/or star-forming efficiency and/or escape fraction.
The value of $N_{\rm ion}$ can be almost 40 times larger than its
value at $z<6$. Also note that $N_{\rm ion}$ need not be a monotonic function
of $z$. For example, the mean model, which is constant
for $z<6$, shows an increase for $z > 6$ followed
by a decrease at $z \approx 7$. The plot shows a subsequent increase around 
$z \approx 11$, however one should remember that the information 
contained within eigenmodes are severely limited at these epochs.

From the plot of $\Gamma_{\rm PI}(z)$ (top-middle panel), we find that the 
mean model is consistent with the observational data at $z < 6$, as expected. 
The errors corresponding to 95\% confidence limits are also smaller at
$z < 6$ for reasons discussed above. The photoionization rate
for the fiducial model shows a 
smooth rise at $z > 6$ with a peak around $z \approx 10$, however model 
described by the mean values of the parameters shows a 
much sharper rise and much prominent peak. The location
of the peak is around $z \sim 6.5$. The highest value of $\Gamma_{\rm PI}$ allowed
by the data can be as high as $10^{-10}$ s$^{-1}$ (95\% confidence level), which
is about 100 times the values typically observed at $z < 6$. 
The prominent peak-like structure is also present in the 
$\de N_{\rm LL}/\de z$ (top-right panel). Interestingly, the high-$\Gamma_{\rm PI}$
models predict that $\de N_{\rm LL}/\de z \approx 0$ at $z \sim 6.5$, hence
any sighting of LLS at these epochs would put more constraints on the models.

The limits on $\tau_{\rm el}$ (bottom-left panel) are, as expected, similar
to the WMAP7 constraints. We find that the mean 
$\tau_{\rm el}$ is slightly higher than the best-fit WMAP7 value because 
a wide range of models with 
early reionization are allowed by the data. 

The constraints on the reionization history can be seen from
the plot of $Q_{\rm HII}(z)$ (bottom-middle panel). The growth of 
$Q_{\rm HII}$ for the fiducial model is somewhat gradual. On the other 
hand, the mean model, which is characterized by sharp peak structures in 
$N_{\rm ion}$ and $\Gamma_{\rm PI}$ at $z > 6$, shows a much faster rise 
in $Q_{\rm HII}$ at initial stages, though the completion of reionization takes place only at
$z \approx 6$. The shaded regions show that reionization can be complete 
as early as $z \approx 10.5$ (95\% confidence level). These models of early reionization are essentially
characterized by high $N_{\rm ion}$ at $6.5<z<10$ (so that 
enough contribution to $\tau_{\rm el}$ is achieved to match the 
WMAP7 constraints) followed by a sharp decrease 
at $z < 6.5$ so that the emissivity becomes low enough to match 
the photoionization rate obtained from Ly$\alpha$ forest data.

Similar conclusions can be obtained from the plot of $x_{\rm HI}(z)$ 
(bottom-right panel). 
In general, the models allowed by the 95\% confidence limits are consistent
with the available data points (shown by points with error-bars).
Models of very early reionization (i.e., those with 
high $N_{\rm ion}$ at $6.5 < z < 10$) show
sharp decrease in $x_{\rm HI}$ at $z \approx 10$ and it can become as
low as $10^{-6}$ at $z \approx 6.5$. However, the neutral fraction has
to increase sharply again at $z < 6.5$ (corresponding to sharp decrease in 
$N_{\rm ion}$) so as to match the Ly$\alpha$ forest constraints. Thus
the evolution of $x_{\rm HI}$ is not monotonic for these models. On the other hand, models with relatively smoothly evolving $N_{\rm ion}$ (ones similar to the
fiducial model) show gradual decrease in $x_{\rm HI}$ between $6 < z < 10$
and it smoothly matches the Ly$\alpha$ forest data. The evolution of the
neutral fraction is thus monotonic in such models with smoothly evolving
$N_{\rm ion}$.

If we now go back to the lower portion of  
Table \ref{tab:PCA_likelihood}, 
we find that reionization is 50\% complete between redshifts 9.6 -- 12.0 (95\% confidence level), while it is 
almost (99\%) complete between redshifts 5.8 -- 10.6 (95\% confidence level). 
Note that the lower limit on the 
redshift of reionization (5.8) is imposed as a prior
on the parameters.

Thus, the PCA shows that a wide range of reionization histories is still allowed
by the data. Reionization can be quite early or can be gradual and late, depending on the behavior of $N_{\rm ion}(z)$. Hence, if one considers only the data
we have used, it is practically impossible to put any sensible 
constraints on chemical feedback and/or the evolution of star-forming
efficiencies and/or escape fractions. While this might seem somewhat
disappointing at the moment, one can hope for much better constraints
in near future when the magnitude of data sets are going to rise manifold. 
In fact, in order to keep the analysis simple, 
we have not used all the data sets available.
For example, the constraints on the Ly$\alpha$ and Ly$\beta$ 
transmitted fluxes now extend beyond $z = 6$ and possibly could constrain the
models much more. However, our numerical code takes significantly more time
while calculating the transmitted fluxes and also there remain uncertainties in 
the theoretical modelling of the IGM at such redshifts (like the 
distribution of baryonic matter and the scatter in the temperature-density relation); 
hence we have worked simply with the
constraints on $\Gamma_{\rm PI}$.
Similarly the distribution of 
LLS at $z > 6$ could also be important in ruling out some of the
allowed models. At present, there exists a data point at $z \approx 6$ which 
put limits $\de N_{\rm LL}/\de z = 8.91 \pm 3.49$. On the other hand, very high
emissivity models predict $\de N_{\rm LL}/\de z \approx 0$ at $z \sim 6.5$. 
Hence constraints on LLS distribution at 
$z\sim 6.5$ can be helpful in shrinking the allowed parameter space 
significantly.

We should also mention that the constraints obtained through the PCA are widely
different from those obtained using the chemical feedback model 
of Section \ref{sec:cfmodel} involving PopII and PopIII stars. The model in
Section \ref{sec:cfmodel} uses a particular prescription for chemical feedback 
and assumes constant $N_{\rm ion, II}, N_{\rm ion, III}$, which results in an
effective $N_{\rm ion}(z)$ which is smoothly evolving and monotonically
increasing with $z$. On the other hand, the models allowed by the PCA do not
have any physical constraint regarding how $N_{\rm ion}$ should evolve. 
It turns out that in absence of any physical motivation, current data does allow
for non-monotonic $N_{\rm ion}$ which may contain sharp features. Hence
it is not surprising that the shapes of the allowed models are quite
different from the chemical feedback models.

\section{Discussion and Summary}

In this work, we have used a semi-analytical model 
\cite{2005MNRAS.361..577C,2006MNRAS.371L..55C} to study the observational constraints on reionization.
Assuming that reionization at $z > 6$ is primarily driven by stellar
sources, 
we have developed a formalism based on principal component analysis to model the
unknown function $N_{\rm ion}(z)$, the number of photons in the IGM
per baryon in collapsed objects. We have used three different sets of
data points, namely, the photoionization rates 
$\Gamma_{\rm PI}$ obtained from Ly$\alpha$ forest Gunn-Peterson
optical depth, WMAP7
data on electron scattering optical depth $\tau_{\rm el}$,
and the redshift distribution of
Lyman-limit systems $\de N_{\rm LL}/\de z$ 
at $z \sim 3.5$.

The main findings of our analysis are:

\begin{itemize}

\item The elements of the Fisher information matrix have larger values
for $z < 6$ where most of the data points are. There is hardly
any information at $z > 14$, implying that no information
on star-formation and/or chemical feedback can be obtained at
these redshifts using the available three data sets.

\item To model $N_{\rm ion}(z)$ over the range $2 < z < 14$ 
it is necessary to include 5 modes. Using a larger number of modes 
improves the agreement but at the cost of increasing errors.

\item One may not be able to  recover the actual 
form of $N_{\rm ion}(z)$ using only these 5 modes, however the
recovery of $\Gamma_{\rm PI}$ and $\de N_{\rm LL}/\de z$
at $z < 10$ is quite satisfactory and that of $Q_{\rm HII}, x_{\rm HI}$ 
is excellent.

\item It is not possible to match available reionization data with a constant
$N_{\rm ion}$ over the whole redshift range, i.e. $N_{\rm ion}$ must increase 
at $z > 6$. This is a signature of either of a changing IMF induced by chemical 
feedback and/or evolution in the star-forming efficiency and/or 
photon escape fraction of galaxies. The data allows for non-monotonic 
$N_{\rm ion}(z)$ (and consequently of $x_{\rm HI}$). In particular, 
reionization histories could show sharp features around $z \approx 7$.

\item The PCA implies that reionization must be 99\% completed between 
$5.8 < z < 10.3$ (95\% confidence level)
and is expected to be 50\% complete at $z \approx $ 9.5--12.

\end{itemize}

Our analysis provides the widest possible range in reionization histories (shown in Fig. 7) allowed by available data sets. It is, in some sense, unfortunate that there still exists a wide range of reionization scenarios that are allowed by the data. While the constraints 
at $z < 6$ are quite tight, one requires additional data points at
$z > 6$ to improve constraints on models of feedback and reionization. 
The most obvious addition would, of course, be observation of Gunn-Peterson trough in
more QSOs at higher redshifts. In parallel, it is expected that observations of GRBs and 
Ly$\alpha$ emitters could constrain $x_{\rm HI}$ at $z > 6$, which again would result in improved constraints. Finally, observations of large-scale EE polarization signal by
future CMB probes, like {\it Planck}\footnote{http://www.esa.int/SPECIALS/Planck/index.html}, would be extremely important in probing the evolution of $N_{\rm ion}$ at $z > 6$. Since the constraints obtained from the data are still unsatisfactory,
there remains ample scope for developing physically-motivated theoretical models which 
can match a wide-variety of available data. This, in turn, requires significant improvement in our understanding of processes like chemical feedback and also the evolution of star-forming efficiencies and escape fraction.

\section*{Acknowledgements}

We would like to thank Dhiraj Kumar Hazra and Atri Bhattacharya for their help and suggestions regarding numerical computations. Computational
work for this study was carried out at the cluster computing facility
in the Harish-Chandra Research Institute
\footnote{http://cluster.hri.res.in/index.html}.

\bibliography{mnrasmnemonic,IGM-ADS}
\bibliographystyle{mnras}

\end{document}